\documentstyle[twoside,fleqn,espcrc2]{article}

\newcommand{\AmS}{{\protect\the\textfont2
  A\kern-.1667em\lower.5ex\hbox{M}\kern-.125emS}}
\title{Irreversibility temperature from magnetic relaxation}

\author{Mahesh Chandran\address{Department of Condensed Matter Physics and Material Sciences, \\ 
        Tata Institute of Fundamental Research, Colaba, Mumbai 400 005.}}
       
\begin{document}

\begin{abstract}
Simulation of the magnetic relaxation in a model of hard superconductor reveals a new time scale below the irreversibility temperature. The relaxation in this new time scale, which appears in an intermediate time window, is a power law and is related to self organization of the magnetic flux around the critical current density $J_c$ during the relaxation. Emergence of a new time scale which along with the time scale for long time relaxation due to thermally activated process dynamically identifies the irreversibility temperature.
\vspace{1pc}
\end{abstract}

\maketitle

\section{Introduction}
Magnetic relaxation in high-$T_c$ superconductors is experimentally known to be non-logarithmic over a large time interval \cite{yesh}. 
Though the non-logarithmic behaviour is related to a diverging pinning potential within vortex glass theory \cite{malz} and collective creep theory \cite{feigl}, the details of the experimental behaviour is not consistent with the theoretical predictions. Specifically, the normalized relaxation rate is found to be temperature independent with a value that cannot be accounted by the theory \cite{yesh}. Moreover, no clear picture exists for the change in relaxation behaviour across the irreversibility temperature $T_{irr}$ and into a true superconducting state with finite persistent current.

\section{The model}
In order to understand generic processes involved during the relaxation in hard type-II superconductors, we have simulated relaxation of the thermoremanent magnetisation in a 2D Josephson junction array at finite temperatures. As have been shown previously \cite{mah1}, a 2D Josephson junction array (JJA) with finite screening current is a realistic model for magnetic behaviour of hard superconductors. We consider a 2D JJA of size $N\times N$ in a transverse magnetic field. The equation of motion for the gauge invariant phase difference across a junction $\phi$ is given by
\begin{eqnarray}
\frac{d\phi}{d\tau} & \;= & \;{\sf M}^{T}{I}_{m}\;-\; \sin \phi\;\;+\;\; X(\tau),
\nonumber \\
{\sf M}\phi & \;= & \;-2\pi f\;-\;\frac{1}{\lambda^{2}_{J}}{I}_{m}.
\end{eqnarray}
Here, the current in a cell $I_{m}$ is scaled by the critical current $I_{c}$ of the junction. The $\tau=\frac{2\pi RI_{c} }{\Phi_{0}}t$ represents dimensionless time, and $\lambda^{2}_{J}=\frac{\Phi_{0}}{2\pi L_{0}I_{c}}$ is the dimensionless penetration depth analogous to London's penetration depth of a bulk superconductor ($\Phi_{0}$ is a quantum of flux, $L_0$ is the self-inductance of a unit cell, and $R$ is the normal state resistance of the junction). The applied magnetic flux is represented by $f=\Phi_{ext}/\Phi_{0}$. The matrix ${\sf M}$ is directed loop-sum operator and is equivalent to lattice curl operation. The temperature bath is attached through the noise term $X(\tau)$ with $\langle X_{\bf r}(t) \rangle\!=\!0$ and $\langle X_{\bf r}(\tau)X_{\bf r'}(\tau') \rangle = 2T\delta(\tau-\tau')\delta_{\bf r,r'}$. The temperature is in units of $I_{c}\Phi_{0}/2\pi k_{B}$. We employed free-end boundary conditions for all the variables. Further details regarding the simulation and setting up of the equation can be found in ref. \cite{mah2}.

\section{Results and discussions}
We present here the results for an array of size $N=16$. The Fig.\ref{fig1} shows the relaxation over 6 decades in $\tau$ for some selected temperatures. The relaxation behaviour changes markedly across two temperatures : $T_{cr}\approx 0.24$ and $T_{sc}\approx 0.04$. At $T_{cr}$, the relaxation curve develops a kink in an intermediate time interval which at lower temperature develops into a plateau on which the dynamics is almost frozen. The width of this plateau thus sets a new time scale $\tau_{\beta}$ for the relaxation of the magnetisation. Experimentally, the $T_{irr}$ is obtained as the temperature at which the field cooled (FC) and zero field cooled (ZFC) susceptibility $\chi(T)$ differs. Inset of Fig.\ref{fig1} shows $\chi(T)$ for the model. The temperature $T_{cr}$ at which the kink appears in the relaxation curve is also the temperature at which the $\mid\chi_{FC}-\chi_{ZFC}\mid >0$. This allow us to conclude that $T_{cr}$ is the irreversibility temperature appearing in the relaxation. This is also consistent with the previous simulation study in which magnetic irreversibility is found to set below this temperature \cite{mah3}. For $T_{sc}<T<T_{cr}$, the magnetisation $M(\tau\rightarrow\infty)=0$. At $T_{sc}$, the relaxation is frozen on the plateau as $\tau\rightarrow\infty$, thus establishing a finite persistent current density (hence, remanent magnetisation). We consider this as the transition into the true superconducting state. 
By explicitly observing the flux distribution, we find that the plateau is related to the crossover from a supercritical state with $J>J_{c}$ to a subcritical state with $J<J_{c}$. We attribute this crossover as arising due to self-organization during the relaxation process. The $\tau_{\beta}$ increases rapidly with decreasing $T$.

For $T_{sc}<T$, the long time relaxation fits to $\exp[-(\tau/\tau_{\alpha})^{\alpha}]$. The characteristic time scale $\tau_{\alpha}$ increases rapidly below $T_{cr}$. One also observes $\log(\tau/\tau_{0})$ behaviour for 1-2 decades far from the plateau. This is the regime in which thermal activation is a dominant process and can be used to extract the characteristic time scale $\tau_{0}$ for relaxation at long time. Fig.\ref{fig1} inset (b) shows the temperature dependence of $\tau_{0}^{-1}\propto$ (magnetic diffusivity) and is plotted as a function of $T-T_{sc}$. Though $\tau_{0}$ follows the Arrhenius law, a power law at low temperatures as indicated by the straight line in the plot is surprising. The $T_{sc}$ obtained from fitting the time scale for long time relaxation matches well with the $T$ at which $M(\tau\rightarrow\infty)=M_{0}> 0$. The $T_{cr}$ appears as the temperature at which $\tau_{0}^{-1}$ deviates from the power law as marked in Fig. 1 inset(b).

In conclusion, we have shown that a new time scale govern the flux dynamics below $T_{irr}$ which arises due to self-organization of the magnetic flux during the relaxation. This conclusively proves that $T_{irr}$ is associated with the crossover in dynamical behaviour for magnetic flux in type-II superconductors.

{\em Figure Caption}
The normalized magnetisation $M(\tau)/M(0)$ for some selected temperatures (marked along the curves) for $f=5$ and $N=16$. Inset: (a)shows the $\chi(T)$ for FC and ZFC conditions ($T_{irr}$ is marked on the curve). (b) The $\tau_{0}^{-1}$ and $\tau_{\beta}$ as a function of $T-T_{sc}$ on log-log plot. The vertical axis label indicates the exponent only.

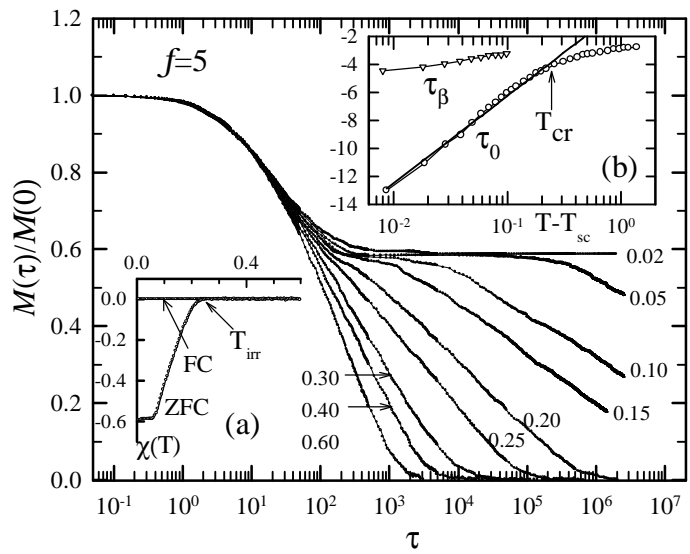
\end{document}